\documentclass[usenatbib, twocolumn, nofootinbib, reprint,prd]{revtex4-1}

\usepackage{newtxtext,newtxmath}
\usepackage[normalem]{ulem}
\usepackage{multirow}
\usepackage{booktabs}
\usepackage{color}
\usepackage{amsmath,amsfonts,bm}
\usepackage{hyperref}

\usepackage[T1]{fontenc}
\usepackage{ae,aecompl}

\usepackage{graphicx}	
\usepackage{amsmath}	
\usepackage{amssymb}

\newcommand{\fittingradius}{10^{\prime}}

\newcommand{\mvir}{M$_{200_{\rm crit}}$}
\newcommand{\arcmin}{^{\prime}}

\newcommand{\snronehalotermonly}{10.7}
\newcommand{\stacksigmawithtwohaloterm}{17}

\newcommand{\binonemassnotwohaloterm}{6.45}
\newcommand{\bintwomassnotwohaloterm}{11.88}
\newcommand{\binthreemassnotwohaloterm}{28.55}
\newcommand{\stackmassnotwohaloterm}{10.80}

\newcommand{\binonemasswithtwohaloterm}{4.18}
\newcommand{\bintwomasswithtwohaloterm}{6.93}
\newcommand{\binthreemasswithtwohaloterm}{18.84}
\newcommand{\stackmasswithtwohaloterm}{6.25}

\newcommand{\binonemassnotwohalotermerror}{0.8}
\newcommand{\bintwomassnotwohalotermerror}{0.9}
\newcommand{\binthreemassnotwohalotermerror}{1.2}
\newcommand{\stackmassnotwohalotermerror}{0.6}

\newcommand{\stackmasswithtwohalotermlowdelz}{6.1}

\newcommand{\stackmasswithtwohalotermhighdelz}{5.9}

\newcommand{\binonemasswithtwohalotermlowdelzwithcorr}{4.13}
\newcommand{\bintwomasswithtwohalotermlowdelzwithcorr}{7.13}
\newcommand{\binthreemasswithtwohalotermlowdelzwithcorr}{18.84}

\newcommand{\binonemasswithtwohalotermhighdelzwithcorr}{3.98}
\newcommand{\bintwomasswithtwohalotermhighdelzwithcorr}{6.62}
\newcommand{\binthreemasswithtwohalotermhighdelzwithcorr}{18.44}

\newcommand{\WxS}{WISE $\times$ SCOS} 
\newcommand{\msol}{$\mbox{M}_{\odot}$}

\newcommand{\shm}{$\mbox{M}_{*}-\mbox{M}_{h}$}

\begin{document}

\title{Imprints of gravitational lensing in the {\it Planck} CMB data at the location of WISExSCOS galaxies}
\author{Srinivasan Raghunathan}\email{srinivasan.raghunathan@unimelb.edu.au}
\author{Federico Bianchini}\email{federico.bianchini@unimelb.edu.au}
\author{Christian L. Reichardt}\email{christian.reichardt@unimelb.edu.au}
\affiliation{School of Physics, University of Melbourne, Parkville, VIC 3010, Australia}

\keywords{CMB lensing -- keyword2 -- keyword3}

\date{Accepted XXX. Received YYY; in original form ZZZ}

\begin{abstract}
We detect weak gravitational lensing of the cosmic microwave background (CMB) at the location of the WISE $\times$ SuperCOSMOS (\WxS) galaxies using the publicly available {\it Planck} lensing convergence map. By stacking the lensing convergence map at the position of 12.4 million galaxies in the redshift range $0.1\le z \le 0.345$, we find the average mass of the galaxies to be \mvir = \stackmasswithtwohaloterm\ $\pm$ \stackmassnotwohalotermerror\ $\times\ \mbox{10}^{12}\ \mbox{M}_{\odot}$. The null hypothesis of no-lensing is rejected at a significance of $\stacksigmawithtwohaloterm \sigma$. We split the galaxy sample into three redshift slices each containing $\sim$4.1 million objects and obtain lensing masses in each slice of \binonemasswithtwohaloterm\ $\pm$ \binonemassnotwohalotermerror, \bintwomasswithtwohaloterm\ $\pm$ \bintwomassnotwohalotermerror, and \binthreemasswithtwohaloterm\ $\pm$ \binthreemassnotwohalotermerror\ $\times 10^{12}\ \mbox{M}_{\odot}$. 
Our results suggest a redshift evolution of the galaxy sample masses but this apparent increase might be due to the preferential selection of intrinsically luminous sources at high redshifts. 
The recovered mass of the stacked sample is reduced by 28\% when we remove the galaxies in the vicinity of galaxy clusters with mass \mvir = $2 \times 10^{14}\ \mbox{M}_{\odot}$.
We forecast that upcoming CMB surveys can achieve 5\% galaxy mass constraints over sets of 12.4 million galaxies with \mbox{\mvir = $1 \times 10^{12}\ \mbox{M}_{\odot}$} at $z=1$.

\end{abstract}

\maketitle



\section{Introduction}
The path of cosmic microwave background (CMB) photons is perturbed by intervening dark matter haloes and associated structures between the observer and the last scattering surface. The magnitude of the deflection $\vec{\alpha}(\hat{n})$ is directly proportional to the gradient of the underlying lensing potential $\phi$ and provides an accurate mapping of the total matter distribution in the Universe. Several previous studies have detected lensing of the CMB due to the large-scale structure (LSS) both in CMB temperature and polarization maps \citep[~and references therein]{pb2014, story2015, abcde, b2k2016, sherwin2017} and by cross-correlating the CMB lensing maps against biased tracers of the matter field such as galaxies \citep{smith2007, bianchini15, sptdes, hill16, ferraro16, siyu2017}.
The lensing on arcminute scales due to massive dark matter haloes has also been detected by stacking techniques \citep{act_cmass2015, baxter2015, planckXXIV2015, geach2017, baxter2017}. This small-scale CMB-halo lensing is especially interesting as it allows us to accurately measure the masses of the astrophysical objects \citep{seljak2000, vale2004, dodelson2004, holder2004, lewis2006, maturi2005, hu2007, yoo2008, melin2015}. The method is more powerful than galaxy lensing at high redshifts where the observed source galaxy counts drop significantly, degrading the lensing signal-to-noise (S/N). At low redshifts, it is complementary to galaxy lensing allowing systematic checks. The current mass estimates using CMB-halo lensing are uncertain at $\ge$20\% level \citep{baxter2017}, with the error budget being dominated by statistical uncertainties. However, the field is rapidly evolving with improved CMB maps expected from the upcoming CMB surveys \citep{benson2015_3g, advact_2016, cmbs4_2016}.

An accurate mass measurement of the largest dark matter haloes (\mvir $\gtrsim \mbox{10}^{13.5}$ \msol) is important for cosmology as these haloes are signposts of the highest density peaks in the Universe and their abundance as a function of mass and redshift is a sensitive probe of structure growth in the Universe \citep{allen2011}. Obtaining accurate masses of intermediate and lower mass haloes is also important to understand the effects of baryon physics on the formation and evolution of galaxies \citep{mo2010}. For example, determining the stellar-to-halo (\shm) mass relation and its redshift evolution can give detailed insights on the history of star formation \citep[~and references therein]{leauthaud2012}. 
Although CMB-halo lensing signal is faint for a single object (S/N < 0.1 for \mvir=1 $\times \mbox{10}^{12}\ \mbox{M}_{\odot}$ at CMB-S4 noise levels), the method offers excellent prospects in determining \shm\ relation out to very high redshifts for the upcoming LSS surveys expected to detect billions of galaxies.

In this work, we extract the halo lensing signal using the {\it Planck} data and the all-sky WISE $\times$ SuperCOSMOS (\WxS\ hereafter) galaxy catalogue \citep{bilicki:2016}. Specifically, we stack the {\it Planck} convergence $\bm{\kappa}$ map at the positions of 12.4 million (M) \WxS\ galaxies. 
The convergence map quantifies the amount of magnification/demagnification of the source, CMB anisotropies in this case, at a given location and is related to the underlying lensing potential $\phi$ as $\bm{\kappa}= -\frac{1}{2}\nabla^{2}\phi$.
We also perform a tomographic stacking analysis by splitting the \WxS\ sample in three redshift bins. Throughout this work, we use the $\Lambda$CDM cosmology obtained from the chain that combines {\it Planck} 2015 data with external datasets \texttt{TT,TE,EE+lowP+lensing+ext} \citep{planckcosmo2015}. We define all the halo quantities with respect to the radius $R_{200}$ defined as the region within which the average mass density is 200 times the critical density of the universe at the halo redshift.

\begin{figure}
\centering
\hspace{-3mm}
\includegraphics[width=\columnwidth, keepaspectratio]{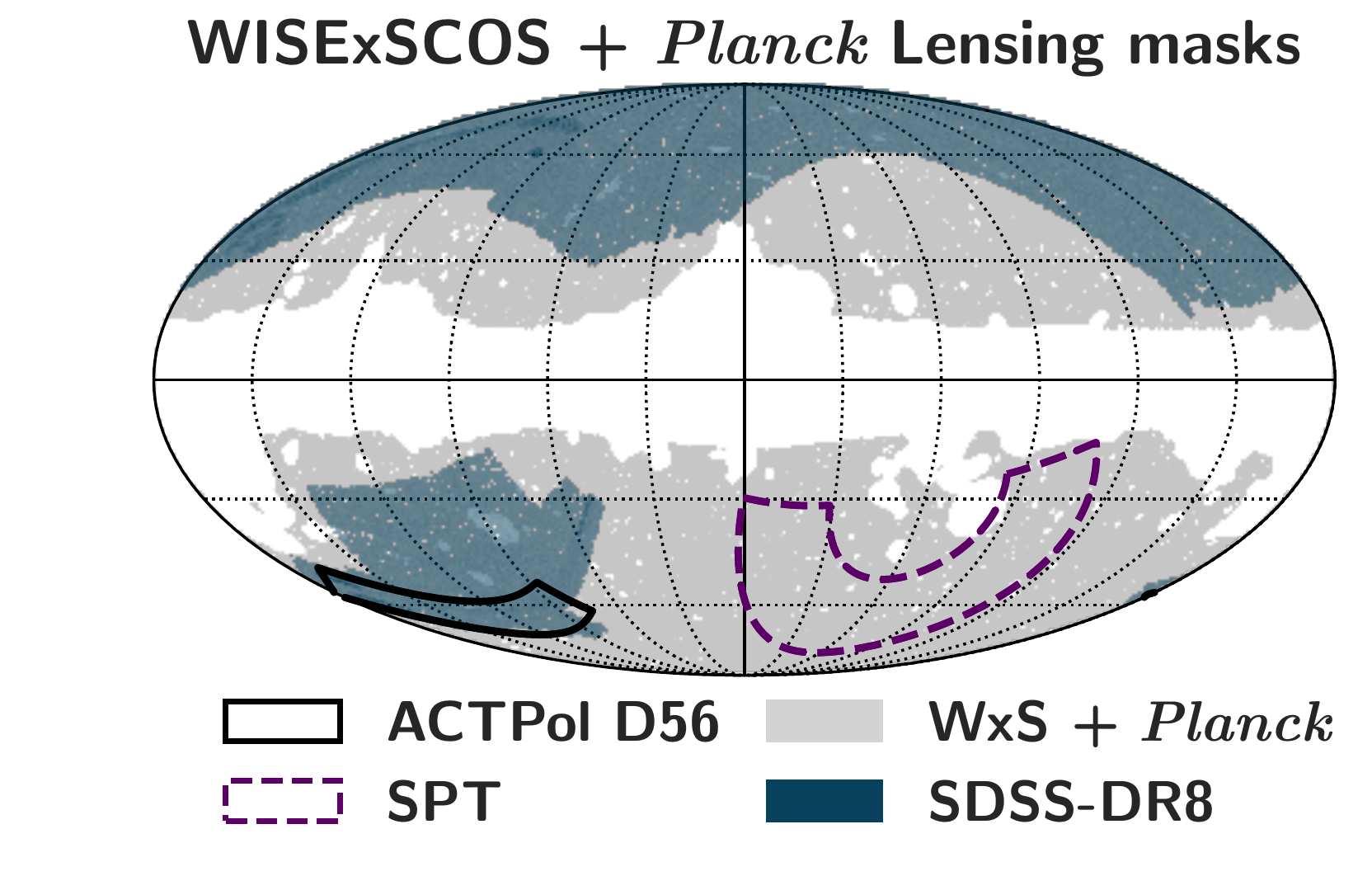}
\caption{Survey footprints in galactic coordinates. The lighter-grey shaded area represents the region of the sky used in our analysis that survives the combined \textit{Planck} lensing and \WxS\  mask ($f_{\rm sky} = 0.65$). The darker bluish region denotes the Sloan Digital Sky Survey (SDSS)-Data Release 8 (DR8) footprint over which the \texttt{redMaPPer} cluster catalog is provided, while the black-solid and magenta-dashed lines indicate the ACTPol D56 \citep{naess2014} and SPT-SZ \citep{george2015} survey area.}
\label{fig_survey_footprints}
\end{figure}

\section{Datasets}
\label{sec:datasets}
In this work, we use the the publicly available\footnote{\url{http://pla.esac.esa.int/pla/}.} 2015 {\it Planck} lensing convergence map \citep{abcde} and the \WxS\footnote{\url{http://ssa.roe.ac.uk/WISExSCOS.html}.} galaxy catalogue. 
The galaxy catalogue \citep{bilicki:2016} contains 20.5M objects and was constructed by cross-matching two of the largest all-sky galaxy samples, the mid-IR AllWISE \citep{cutri:2013} and the optical SuperCOSMOS \citep{peacock:2016}. Both the {\it Planck} lensing map and the \WxS\ catalogue were accompanied by masks intended to remove contaminated regions, like the regions of high-galactic emission. We create a combined mask from {\it Planck} and \WxS\ masks, shown as the light-shaded grey region in Fig. \ref{fig_survey_footprints}. The galaxy sample after employing this mask has photometric redshifts (photo-$z$s) distributed between $0 \lesssim z \lesssim 0.4$ ($\tilde{z}\simeq 0.2$) with a normalized scatter $\sigma_z=0.033$. From this sample, we remove objects outside the redshift range $0.1 \le z \le 0.345$ as the stellar contamination outside this range is greater than 20\% in \WxS\ catalogue \citep{bilicki:2016}. Our final sample contains 12.4M objects with $f_{\rm sky} = 0.65$.

\begin{figure*}
	\includegraphics[width=1.\textwidth, keepaspectratio]{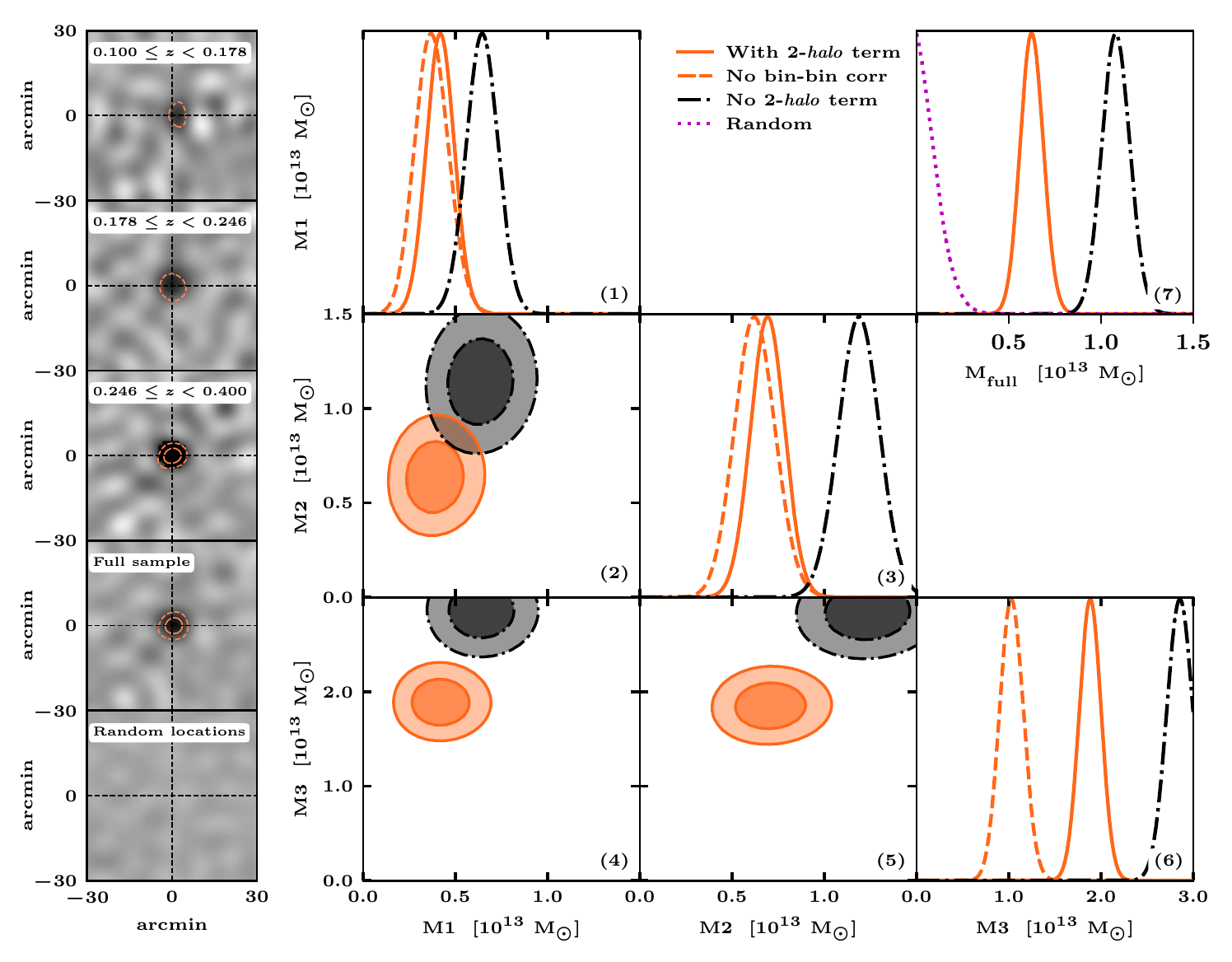}
    \caption{\textit{Left panels:} Stacked \textit{Planck} convergence maps at the location of \WxS\ galaxies in the three redshift slices and in the full sample (upper four boxes), as well as at a number of random positions equal to that of the full galaxy sample (a null test). The dashed and solid contours correspond to regions above 3 and $5\sigma$ respectively.
The colour scale has been restricted to the range $[-0.002,0.0035]$ for all panels.
\textit{Right panels:} Panels 1-6 show the lensing masses obtained in the three redshift bins when correlations between different bins are included in the fitting. The contours in the off-diagonal panels show the $1\sigma, 2\sigma$ constraints. The orange-soild and black-dash-dotted curves correspond to constraints obtained with and without the inclusion of the {\it 2-halo term}. The orange-dashed curves on the diagonal panels are the lensing mass for the three bins when the bin-bin correlations are ignored during fitting. The lensing mass of the full sample along with the null test (purple-dotted) is shown in panel 7.}    
    \label{fig1_stack_likelihoods}
\end{figure*}

\section{Methods}
\label{sec:methods}
The aim of this work is to measure the average mass of the dark matter haloes that host the \WxS\ galaxies in a tomographic approach by stacking the \textit{Planck} CMB convergence map.
To achieve this, we extract $60^{\prime} \times 60^{\prime}$ cutouts 
from the {\it Planck} $\bm{\kappa}$ map
at the location of each \WxS\ galaxy by projecting the full-sky map to a tangential flat-sky projection at $1^{\prime}$ resolution using \texttt{healpy.gnomview} command.  
Note that the estimate of the background CMB gradient with a quadratic estimator is underestimated at the location of a dark matter halo due to the additional lensing caused by the halo. This results in a small bias in the reconstructed lensing signal \citep{maturi2005, hu2007} and can be mitigated by estimating the gradient at scales larger than $L\le2000$ \citep{hu2007}. Ignoring this low-pass filter, as is the case here when using {\it Planck} lensing convergence maps, is not an optimal method of extracting the CMB-halo lensing signal. 
However, this causes negligible effect to the analysis presented here because of the high noise levels in the {\it Planck} maps.
For the tomographic slicing, we split the sample into three redshift bins\footnote{Given the photo-$z$ uncertainties we must choose redshift bins of width $\Delta z \gtrsim 0.04$.} comprising $\sim$4.1M objects each: $0.1 \le z < 0.178$, $0.178 \le z < 0.246$, and $0.246 \le z \le 0.345$ and stack, i.e. average the respective number of cutouts in each bin to obtain the final cutout. To validate the analysis we also stack the masked $\bm{\kappa}$ map at a number of random sky locations equal to that of our galaxy sample. These stacked cutouts are shown in the left panel of Fig. \ref{fig1_stack_likelihoods}. 

We model the halo density profiles using Navarro-Frenk-White (NFW) dark matter (DM) profile \citep{nfw1996} and obtain $\bm{\kappa}^{\rm NFW}$ using Eq. (2.8) of \citet{bartelmann1996}.
We also include the lensing due to correlated structures along the line-of-sight ({\it 2-halo term}) \citep{seljak2000b,cooray2002} using Eq. (13) of \citet{oguri2011}. We adopt the \citet{tinker:2010} formalism to calculate the bias $b_h(M,z)$ for a halo with mass $M \equiv M_{200_{\rm crit}}$. Thus, our model for the total lensing convergence is $\bm{\kappa}^{\rm m}(\bm{\theta}) = \bm{\kappa}^{\rm 1h} + \bm{\kappa}^{\rm 2h}$. 
Finally, we filter out modes $L > 2048$ in $\bm{\kappa}^{\rm m}(\bm{\theta})$ to match the filtering adopted in the {\it Planck} lensing map.

We determine the average best-fit galaxy halo mass $M_{200_{\rm crit}}$ of the stacked sample in each redshift bin by maximizing the likelihood
\begin{equation}
\label{eq:likelihood}
-2\ln\mathcal{L} = \sum_{\theta, \theta^{\prime} = 0}^{10^{\prime}}\left[\hat{\kappa}(\theta) - \hat{\kappa}^{\rm m}(\theta)\right] \hat{\bf{C}}_{\theta, \theta^{\prime}}^{-1} \left[\hat{\kappa}(\theta^{\prime}) - \hat{\kappa}^{\rm m}(\theta^{\prime})\right].
\end{equation}
Here $\hat{\kappa}(\theta)$ and $\hat{\kappa}^{m}(\theta)$ are the concatenated radially binned profiles of the data and model in the three redshift bins up to $\fittingradius$ with $\Delta\theta = 2.^{\prime}5$. The results are stable to our choice of fitting radius and change only marginally when we increase the radius to $12.^{\prime}5$.
We use the median redshift $\tilde{z}_{i}$ of the galaxies in each bin $i$ when determining $\bm{\kappa}^{\rm m}$ for different halo masses. We use the \citet{duffy:2008} mass-concentration relation to calculate the concentration parameter $c_{200}(M, \tilde{z}_{i})$ for the NFW haloes. 
The covariance matrix $\hat{\bf{C}}_{12 \times 12}$ is calculated using jackknife resampling by dividing the survey region into $N=1000$ samples using

\begin{equation}
\label{eq:likelihood}
\hat{\bf{C}}_{\theta, \theta^{\prime}} = \frac{N-1}{N} \sum\limits_{j=1}^{N} \left[\hat{\kappa}_{j}(\theta) - \left<\hat{\kappa}(\theta)\right>\right] \left[\hat{\kappa}_{j}(\theta^{\prime}) - \left<\hat{\kappa}(\theta^{\prime})\right>\right],
\end{equation}
where $\hat{\kappa}_{j}(\theta)$ is the concatenated data vector from the three redshift bins for the $j^{\rm th}$ jackknife sample. 
We have examined how the eigenmodes of this covariance matrix change when varying the number of jackknife splits, and find the eigenmodes are stable beyond 300 splits. 
It is important to consider the correlations between redshift bins since a given location on sky can be lensed by galaxies in more than one redshift bin.
We also perform the fitting in the three redshift bins without including the correlation between different bins. 
The covariance matrix $\hat{\bf{C}}_{4 \times 4}$ in this case is also obtained using jackknife resampling.
The average mass of the full sample was then determined by combining the likelihoods of the individual redshift bins.
We calculate the $1\sigma$ mass uncertainty $\Delta$\mvir as the point when $\Delta \chi^2 = -2\ln \mathcal{L}(\mbox{M}_{\rm fit}) + 2\ln \mathcal{L}(\mbox{\mvir})$ becomes unity.
The significance of obtaining the estimated lensing mass M$_{\rm fit}$ with respect to no-lensing M$_{\rm fit}$ = 0 is defined as $\sqrt{2 \ln \mathcal{L}(\mbox{\mvir} = \mbox{M}_{\rm fit}) - 2 \ln \mathcal{L}(\mbox{\mvir} = 0)}$.

\begin{table}[]
\caption{Inferred lensing masses for the \WxS\ galaxy sample considering the correlations between different bins.}
\label{tab_lensing_masses_redshift}
\resizebox{0.9\columnwidth}{!}{
\begin{tabular}{|c|c|c|c|}
\hline
 \multirow{2}{*}{Bin} & \multicolumn{3}{c|}{Lensing mass \mvir {[}10$^{12}$ M$_{\odot}${]}}\\
 \cline{2-4}
 & No {\it 2-halo term} & With {\it 2-halo term} & $\sigma$(\mvir)\\ 
\hline\hline

0.100 $\le z$ < 0.178 & \binonemassnotwohaloterm & \binonemasswithtwohaloterm  & $\pm$ \binonemassnotwohalotermerror \\[1ex]\hline
0.178 $\le z$ < 0.246 & \bintwomassnotwohaloterm & \bintwomasswithtwohaloterm & $\pm$ \bintwomassnotwohalotermerror \\[1ex]\hline
0.246 $\le z$ < 0.345 & \binthreemassnotwohaloterm & \binthreemasswithtwohaloterm & $\pm$ \binthreemassnotwohalotermerror \\ [1ex]\hline
\hline
Full sample & \stackmassnotwohaloterm & \stackmasswithtwohaloterm & $\pm$\stackmassnotwohalotermerror \\ [1ex]\hline
%
\end{tabular}
}
\end{table}

\section{Results}
\label{sec:results}

The reconstructed galaxy lensing masses for the three redshift bins are shown in the right panels of Fig. \ref{fig1_stack_likelihoods}. The diagonal panels show the marginalized likelihoods and the off-diagonal contours represent the $1\sigma, 2\sigma$ mass constraints. The orange-solid and the black dash-dotted curves correspond to the fitting performed with and without the {\it 2-halo} term. For reference, the lensing masses obtained without including the correlations between the redshift bins are shown as orange-dashed curves. The lensing masses for the full sample along with the null test (purple-dotted) are shown in the panel 7. 
The no-lensing hypothesis of $\mbox{M}_{\rm fit} = 0$ is rejected at $\stacksigmawithtwohaloterm \sigma$ for the full sample. 
The best-fit lensing masses are tabulated in Table \ref{tab_lensing_masses_redshift}. Since structures correlated with the individual haloes add to the lensing signal, the inferred masses decrease slightly as expected when including the {\it 2-halo term}. The lensing mass of the full sample is \stackmasswithtwohaloterm\ (\stackmassnotwohaloterm) $\times\ \mbox{10}^{12} \mbox{M}_{\odot}$ with (without) the {\it 2-halo term} included. 
Given that the inclusion of the {\it2-halo term} has a significant effect on the fitting, we estimate the signal-to-noise due to $ \bm{\kappa}^{\rm 1h}$ alone by removing the $\bm{\kappa}^{\rm 2h}$ corresponding to the best-fit mass of \stackmasswithtwohaloterm\ $\times\ \mbox{10}^{12} \mbox{M}_{\odot}$ at the median redshift $\tilde{z} = 0.21$. The detection significance of the resultant signal due to $ \bm{\kappa}^{\rm 1h}$ alone is \snronehalotermonly $\sigma$. 
This S/N can be compared with other similar works in the literature namely \citet{act_cmass2015, baxter2015, planckXXIV2015, geach2017, baxter2017}. 
Our results agree well with the expectations from simulations. We obtain $14.4\sigma$ detection significance and $\sigma(\mbox{\mvir}) = 0.9\times \mbox{10}^{12}\ \mbox{M}_{\odot}$
for a sample containing 12.4M galaxies with \mvir = $8 \times \mbox{10}^{12}\ \mbox{M}_{\odot}$ at $z = 0.2$. For the null stack, the lensing mass is consistent with zero - corresponding to a no lensing probability-to-exceed (PTE) value of 0.46. 

We find that the reconstructed halo lensing masses increase with redshift. 
This increase could be due to an evolution of the galaxy properties in the \WxS\ sample. Alternatively, the effect can also be explained due to preferential selection of intrinsically luminous sources at high redshifts given that the \WxS\ is a flux-limited catalogue. Note that the luminous galaxies are generally expected to reside in massive DM haloes \citep{cooray2002}. A thorough investigation of this mass disagreement is left for the future work.

\subsection{Validation of results}

We test the robustness of our results against three effects that could impact our lensing mass measurement: (a) selecting galaxies in dense environments, (b) uncorrelated higher redshift clusters, and (c) redshift binning and uncertainties.


\vspace{2mm}\noindent
\subsubsection{Enhanced lensing signal due to dense environments} 
Given that some of the galaxies reside in galaxy clusters, which at \mvir $\gtrsim 10^{14}\mbox{M}_{\odot}$ can be two or more orders of magnitude more massive than a galaxy, one might worry about what fraction of the stacked lensing signal is due to the host galaxy clusters instead of the galaxy sample. 
We quantity the magnitude of this effect by comparing the recovered masses for all galaxies to the masses when galaxies near clusters are excluded. 
To this end, we use the RM cluster catalog from the SDSS-DR8 (dark-blue region in Fig.~\ref{fig_survey_footprints}) dataset \citep{rykoff2014}. 
The SDSS-RM catalogue contains 14,750 clusters in our region and redshift range of interest ({\it Planck} lensing and \WxS\ masks; $0.1 \le z \le 0.345$). We create RM cluster masks for our three redshift bins. The RM cluster masks remove 1-3\% galaxies in each bin that are within $7^{\prime}$ of a RM cluster.
We stack the convergence before (after) incorporating the RM cluster mask. 
We find that masking galaxies near galaxy clusters ( \mvir  $\gtrsim 2 \times 10^{14} \mbox{M}_{\odot}$) reduces the estimated lensing mass by 30\%, 12\%, 33\% in the three bins and by $\sim$ 28\% for the full sample.
Since the RM catalog does not contain all the clusters in the Universe, a proper way of estimating the environmental effect is to split the \WxS\ catalog into central and satellite galaxies and use a detailed halo occupation distribution (HOD) model to populate the galaxies inside the dark matter haloes \citep{berlind2003}.

\vspace{2mm}\noindent
\subsubsection{Residual foregrounds in {\it Planck} \texttt{SMICA} maps}
The {\it Planck} lensing map was reconstructed using the foreground cleaned \texttt{SMICA} \citep{planck_smica_2015} temperature and polarization maps which contain residual Sunyaev-Zel{'}dovich signals \citep{SZ1970} (both thermal tSZ and kinetic kSZ) and contamination from the cosmic infrared background (CIB) arising primarily due to high-redshift ($z \gtrsim 1$) infrared galaxies \citep{vanEngelen_2014, bobin_2016} that can bias the lensing analysis. 
In this work, however, we work with low-redshift galaxies ($z\le 0.345; \tilde{z} \simeq 0.2$) from \WxS{} catalogue which contribute only a small fraction of the CIB emission \citep{plackciblensing, yu17}. Thus we ignore CIB effects in this analysis.

The SZ effect from the \WxS\ galaxies is expected to be small. However, given that a small fraction of our galaxy sample is close to SDSS RM clusters, we quantify the bias due of SZ signals from RM clusters using the high-resolutions simulations of \citet{sehgal10}. 
We start by simulating LSS lensed CMB of the SDSS footprint using the {\it Planck} 2015 power spectra described earlier.
The simulation was then lensed at the location of 4.5 million \WxS\ galaxies in the SDSS footprint using an halo of mass \mbox{\mvir = 8 $\times 10^{12}$ \msol} at redshift $z = 0.3$ assuming an NFW profile.
We smooth the simulated map using a Gaussian beam of $\theta_{\rm FWHM} = 5'$ and then add instrumental white noise $\Delta T = 45 \mu K'$ corresponding to the {\it Planck} \texttt{SMICA} map.
The simulated map was then passed through a QE pipeline to reconstruct the lensing convergence map and filtered similar to the data by setting the modes $L<8$ and $L > 2048$ to zero.
We refer the reader to \citet{methods_paper_2017} for more details about simulation and the lensing pipeline.
We stack the lensing convergence at the location of \WxS\ galaxies and model them using NFW profile as described in \S\ref{sec:methods}.
For this fiducial case, we find a lensing mass of \mbox{$7.1 \pm 1.9 \times 10^{12}$ \msol} which is less than $0.5\sigma$ away from the true halo mass.

Then, for the same simulated sky above, we add residual SZ emissions from the SDSS RM clusters before performing the halo lensing.
To this end, we pick the tSZ and kSZ signals from \citet{sehgal10} simulations corresponding to RM clusters that survived the masks described in \S\ref{sec:datasets}.
We conservatively assume 100\% residual tSZ signal in the {\it Planck} \texttt{SMICA} temperature map.
We pass them through the lensing pipeline as before and stack the reconstructed convergence map at the location of 4.5 million \mbox{\WxS} galaxies.
In this case, we recover a lensing mass of \mbox{$6.4 \pm 1.8 \times 10^{12}$ \msol} which is $< 0.4\sigma$ from the fiducial case (\mbox{$7.1 \pm 1.9 \times 10^{12}$ \msol}) without the SZ emission. 
Given that the shift due to SZ signals from the massive RM clusters is negligible, we ignore the SZ effect from \WxS\ galaxies which are expected to be two orders of magnitude smaller.
While the effect due to SZ signals is only marginal for this work, care must be taken when using the {\it Planck} lensing map as it reconstructed from \texttt{SMICA} foreground reduced CMB maps which have strong residual SZ contamination at the location of massive clusters \citep{bobin_2016}.

\vspace{2mm}\noindent
\subsubsection{Increased variance due to uncorrelated high-$z$ clusters} 
We also check the increased variance due to presence of higher redshift objects uncorrelated with the galaxy sample. We estimate the change in variance by randomly inserting galaxy clusters in a 25 deg$^{2}$ box behind a galaxy of mass \mvir$= 8.0 \times 10^{12}\ \mbox{M}_{\odot}$. 
For the \citet{tinker:2010} halo mass function (HMF), we find that there should be $\sim$1000 clusters above $z=0.4$ and  \mbox{\mvir $\ge$ 7.5 $\times$ 10$^{13}$ M$_{\odot}$} in 25 deg$^{2}$. 
For simplicity, while we draw the cluster masses from the \citet{tinker:2010} HMF, we assume a spatial Poissonian distribution and fixed redshift of $z=0.7$ for these clusters to add their convergence signal to the simulations. The results show a slight increase in $\sigma$(\mvir) from 18\% to 25\% with only $\le$3\% shift in the inferred mass implying no bias as expected.

\vspace{2mm}\noindent
\subsubsection{Imperfect redshifts} 
We make the simplifying assumption while fitting that all galaxies are at the median redshift of a bin, i.e. $z_{\rm eff} = \tilde{z} = 0.142$, $0.212$, and $0.278$. 
This ignores the finite width of the real galaxy redshift distribution, and any systematic offset in the estimated redshifts. 
We test the redshift sensitivity of the results by shifting $z_{\rm eff}$ up and down to either lower or upper edge of each redshift bin. 
We expect the inferred masses to shift systematically towards higher (lower) masses when the $z_{\rm eff}$ decreases (increases) as the {\it 2-halo term} will correspondingly be under (over) estimated.
However, we find the systematic shifts in mass due to redshift are small compared to the mass uncertainties, and not readily apparent in the results.  
We obtain \binonemasswithtwohalotermlowdelzwithcorr\ (\binonemasswithtwohalotermhighdelzwithcorr), \bintwomasswithtwohalotermlowdelzwithcorr\ (\bintwomasswithtwohalotermhighdelzwithcorr), and \binthreemasswithtwohalotermlowdelzwithcorr\ (\binthreemasswithtwohalotermhighdelzwithcorr) for the three redshift bins and \stackmasswithtwohalotermlowdelz\ (\stackmasswithtwohalotermhighdelz) $\times \mbox{10}^{12}\ \mbox{M}_{\odot}$ for the full sample for $z_{\rm low}$ ($z_{\rm high}$) case.
The differences in the mass estimates for this fairly extreme change in redshift are less than 7\%, well within the error bars.

\subsection{Future forecasts}
Finally, 
we forecast the mass constraints that can be achieved with temperature data from future CMB surveys like Simons Observatory\footnote{\url{https://simonsobservatory.org/}} (SO) and CMB-S4 \citep{cmbs4_2016} that will become operational in the next decade. For this, we simulate CMB-S4 and SO like lensing convergence cutouts assuming an experimental beam of $\theta_{\rm FWHM} = 2\arcmin$. Then we generate lensing reconstruction noise power spectra $N_{L}^{\kappa\kappa}$ 
for modes up to $L = 5000$ assuming temperature map noise levels of $\Delta T$ = 1 and 2.5 $\mu K\arcmin$ respectively. We use $N_{L}^{\kappa\kappa}$ to add noise to the simulations. Considering that the Large Synoptic Survey Telescope is expected to return a galaxy catalogue containing $\mathcal{O}(\gtrsim10^{9})$ sources, enabling a fine splitting based on redshift and galaxy properties, we consider a representative subset at redshift $z = 1.0$ comprising 12.4M galaxies of mass \mvir = $1 \times \mbox{10}^{12}\ \mbox{M}_{\odot}$. 
We find that CMB-S4 (SO) can detect the lensing signal at $93\sigma$ ($82\sigma$) with uncertainty 5.2\% (6.0\%) in the inferred mass.

\section{Conclusions}
\label{sec:conclusions}
In this \textit{Letter}, we stack the \textit{Planck} CMB lensing convergence map around the 12.4M \WxS\ galaxies in the redshift range between $0.1 \le z \le 0.345$ and detect a signal at a significance of $\stacksigmawithtwohaloterm \sigma$. We find a best-fit mass of \mbox{\stackmasswithtwohaloterm\ $\pm$ \stackmassnotwohalotermerror\ $\times \mbox{10}^{12}\ \mbox{M}_{\odot}$} by modelling the lensing signal using NFW profile. We perform a null test by stacking the convergence map at random locations and infer a best-fit lensing mass consistent with zero. Using SDSS-RM clusters, we find that the presence of galaxy clusters in our cutouts can increase the lensing mass by 39\%. 
The uncorrelated higher redshift clusters, as expected, do not bias our results but only increases the mass uncertainty to 25\% from 18\%. The errors introduced due to the use of median redshifts for the stacked sample and the residual foregrounds in the {\it Planck} \texttt{SMICA} maps are less than the statistical uncertainties of the measurements.

Although at an early stage of  development, CMB-halo lensing represents a promising tool for measuring the total mass of astrophysical objects. We forecast that the future CMB stage-4 surveys can achieve 5-6\% mass constraints for high-$z$ galaxies. By piercing the high redshift Universe over large volumes, CMB lensing enables a thorough investigation of the luminous-DM connection in a way complementary to, for example, galaxy weak lensing. This connection is a crucial element not only for a clear understanding of the physics of galaxy formation and evolution, but also for carrying out robust cosmological analyses. 

\subsection*{Acknowledgements}
We thank James Bartlett and Gil Holder for their valuable suggestions and comments. We also thank Benedetta Vulcani for useful discussions. We thank all the three anonymous referees for useful suggestions that helped in shaping this manuscript better. We acknowledge the support from Australian Research Council's Discovery Projects scheme (DP150103208). FB acknowledges support from an Australian Research Council Future Fellowship (FT150100074). We thank the high performance computation centre at University of Melbourne for providing access to the cluster \texttt{spartan.unimelb.edu.au};   University of Chicago for \texttt{spt.uchiago.edu}; and the Wide Field Astronomy Unit (WFAU) at the Institute for Astronomy, Edinburgh, for archiving the \WxS\  catalogue. In this paper we made use of  \texttt{HEALPix}, \texttt{healpy}, and of the \textit{Planck} Legacy Archive (PLA). This research used resources of the National Energy Research Scientific Computing Center (NERSC), a DOE Office of Science User Facility supported by the Office of Science of the U.S. Department of Energy under Contract No. DE-AC02-05CH11231.

\newcommand{\aap}{A\&A}
\newcommand{\procspie}{Proceedings of the SPIE}
\newcommand\mnras{MNRAS}
\newcommand\nata{Nature Astronomy}
\newcommand\apss{Astrophysics and Space Science}

\end{document}